\documentclass{PoS}

\PoS{PoS(LAT2005)221}

\newcommand{\bea}{\begin{eqnarray}}
\newcommand{\eea}{\end{eqnarray}}

\title{$B \to D^* l \nu$ with staggered chiral perturbation theory }

\ShortTitle{$B \to D^* l \nu$ with staggered chiral perturbation
theory }

\author{Jack Laiho$^{a}$\\
        $^a$Fermi National Accelerator Laboratory,
        Batavia, Illinois 60510, USA\\
        Theoretical Physics Department\\
        E-mail: \email{jlaiho@fnal.gov}}

\abstract{An unquenched calculation of the form factor for
$B\rightarrow D^* l \nu$ is needed to improve the determination of
$|V_{cb}|$.  The MILC lattices, computed with a 2+1 improved
staggered action for the light quarks, are well suited to this
purpose. The relevant staggered chiral perturbation theory (SChPT)
must be known in order to correctly account for the ``taste"
breaking discretization effects associated with the staggered
quarks to NLO in $1/ m_{D^*}$. This SChPT calculation is
presented.}

\FullConference{XXIIIrd International Symposium on Lattice Field Theory\\
                 25-30 July 2005\\
                 Trinity College, Dublin, Ireland}

\begin{document}

\section{Introduction}

The CKM element $V_{cb}$ is important for the phenomenology of
flavor physics in determining the apex of the unitarity triangle
in the complex plane.  $|V_{cb}|$ can be determined from inclusive
and exclusive semileptonic $B$ decays, and they are both limited
by theoretical uncertainties.  The inclusive method makes use of
the heavy quark expansion \cite{ball, bigi}, but is limited by the
breakdown of local quark-hadron duality, the errors of which are
difficult to estimate.  The exclusive method requires reducing the
uncertainty of the form factor ${\cal F}_{B\rightarrow D^*}$,
which has been calculated using lattice QCD in the quenched
approximation \cite{hashimoto}. In an effort to eliminate the
errors due to quenching, as well as to reach lighter quark masses,
we anticipate using lattices with 2+1 flavors of light staggered
quarks generated by the MILC collaboration \cite{MILC} to
calculate ${\cal F}_{B\rightarrow D^*}$.  It is therefore
important to understand the taste violations in this heavy-light
system, and for this we use the staggered heavy-light chiral
perturbation theory of Aubin and Bernard \cite{aubin}.

We make use of the result from Aubin and Bernard that at $O(a^2)$
all taste-violations in the Symanzik action are those of the light
quark sector \cite{lee, aubin2}.  The discretization effects due
to the heavy quarks are not taken into account explicitly in the
ChPT, but can be absorbed into the definitions of the lattice
heavy quarks. The heavy-light mesons are combined into a single
field

\bea H_a=\frac{1+ \not \! v}{2}[\gamma^\mu B^*_{\mu a}-\gamma_5
B_a], \eea

\noindent with the conjugate, $\overline{H}_a \equiv \gamma_0
H_a^\dag \gamma_0$.  The light mesons appear in the form $\Sigma =
\sigma^2= \exp [2i\Phi/f]$, where $\Phi$ is a $12\times 12$ matrix
that contains the pions

\bea \Phi = \left(%
\begin{array}{ccc}
  U & \pi^{+} & K^+\\
  \pi^- & D & K^0 \\
  K^- & \overline{K}^0 & S
\end{array}%
\right),\eea

\noindent where $U=\sum_{a=1}^{16} U_a T_a$, \emph{etc}, and
$T_a=\{\xi_5, i\xi_{\mu5}, i\xi_{\mu\nu}, \xi_{\mu}, \xi_I \}$.

The calculation makes use of an expansion in $m_q$ (the light
quark mass), $a^2$, and the heavy-light residual momentum.  The
Lagrangian is

\bea {\cal L}^{(2)} &=& i \textrm{tr}_D[\overline{H}_av^\mu
(\delta_{ab}\partial_\mu + iV^{ba}_\mu)H_b]  +g_\pi
\textrm{tr}_D[\overline{H}_a H_b \gamma^\nu \gamma_5 A^{ba}_\nu] +
{\cal L}_{S\chi PT}, \nonumber \\ && \label{lagrange}\eea

\noindent where $V_{\mu} \equiv (i/2)[\sigma^\dag
\partial_\mu\sigma + \sigma\partial_\mu\sigma^\dag]$, and $A_{\mu} \equiv
(i/2)[\sigma^\dag \partial_\mu\sigma -
\sigma\partial_\mu\sigma^\dag]$.  The leading correction to this
Lagrangian at $1/m_c$ relevant for this calculation is

\bea {\cal L}^{(2)}_M =
\frac{\lambda_2}{m_c}\textrm{tr}_D[\overline{H}_a \sigma^{\mu
\nu}H_a \sigma_{\mu \nu}], \label{mcsplit} \eea

\noindent and gives rise to the splitting between the $D$ and
$D^*$ masses, $\lambda_2 =-\frac{m_c}{8}\Delta^{(c)}
=-\frac{m_c}{8}(m_{D^*}-m_D)$. The staggered light Lagrangian is

\bea {\cal L}_{S}^{(2)} &=& \frac{f^{2}}{8}
    \textrm{tr}[\partial_{\mu}\Sigma\partial^{\mu}\Sigma] +
    \frac{f^{2}B_{0}}{4}\textrm{tr}[\chi^{\dag}\Sigma+\Sigma^{\dag}\chi]
    + \frac{2m^2_0}{3}(U_I+D_I+S_I+...)^2 + a^2{\cal V}, \nonumber \\ && \eea

\noindent where
 \bea  {\cal V}= \sum_k C_k {\cal O}_k + \sum_{k'} C_{k'}
{\cal O}_{k'}, \eea

\noindent are taste breaking operators that can be found in
\cite{aubin2}. The primed operators are taste non-diagonal, while
the unprimed operators are taste-diagonal.  The terms in
Eq.~(\ref{lagrange}) give rise to the NLO chiral logarithms. There
are additional terms not shown in this equation coming from a new
taste-breaking potential that involves both heavy and light
mesons, and although this potential is of higher order than the
terms in Eq.~(\ref{lagrange}), they can contribute analytic terms
at NLO.

\section{Obtaining $|V_{cb}|$}

The differential rate for the semileptonic decay $\overline{B}\to
D^*l\overline{\nu}_l$ is

\bea \frac{d\Gamma}{dw} &=&
\frac{G^2_F}{4\pi^3}m^3_{D^*}(m_B-m_{D^*})^2\sqrt{w^2-1}{\cal
G}(w)|V_{cb}|^2|{\cal F}_{B\rightarrow D^*}(w)|^2 \eea

\noindent where $w=v' \cdot v$ is the velocity transfer from the
initial state to the final state, ${\cal G}(w)$ is a kinematic
factor and ${\cal F}_{B\rightarrow D^*}$ is a matrix element which
must be calculated nonperturbatively. This matrix element is a
combination of several form factors, but at zero recoil it
simplifies to ${\cal F}_{B\rightarrow D^*}(1)=h_{A_1}(1)$.  Heavy
quark symmetry plays an important role in constraining
$h_{A_1}(1)$, leading to the heavy quark expansion \cite{falk,
mannel}

\bea h_{A_1}(1) &=&
\eta_A\left[1-\frac{l_V}{(2m_c)^2}+\frac{2l_A}{2m_c
2m_b}-\frac{l_P}{(2m_b)^2}\right], \nonumber \\ && \label{hA1}\eea

\noindent up to order $1/m_Q^2$. The above works were generalized
to lattice gauge theory in \cite{kronfeld}. The $l$'s are
long-distance matrix elements of the heavy quark effective theory
(HQET).

It was realized that these $l$'s could be computed precisely by
making use of the double ratios of various matrix elements at zero
recoil \cite{hashimoto}:

\bea {\cal R}_+=\frac{\langle D|\overline{c}\gamma_4
b|\overline{B}\rangle \langle \overline{B}|\overline{b}\gamma_4
c|D\rangle}{\langle D|\overline{c}\gamma_4 c|D\rangle \langle
\overline{B}|\overline{b}\gamma_4 b|\overline{B}\rangle} =
\left|h_+(1)\right|^2, \eea \bea {\cal R}_1=\frac{\langle
D^*|\overline{c}\gamma_4 b|\overline{B}^*\rangle \langle
\overline{B}^*|\overline{b}\gamma_4 c|D^*\rangle}{\langle
D^*|\overline{c}\gamma_4 c|D^*\rangle \langle
\overline{B}^*|\overline{b}\gamma_4 b|\overline{B}^*\rangle} =
\left|h_1(1)\right|^2, \eea \bea{\cal R}_{A_1} =\frac{\langle
D^*|\overline{c}\gamma_j \gamma_5 b|\overline{B}\rangle \langle
\overline{B}^*|\overline{b}\gamma_j \gamma_5 c|D\rangle}{\langle
D^*|\overline{c}\gamma_j \gamma_5 c|D\rangle \langle
\overline{B}^*|\overline{b}\gamma_j \gamma_5
b|\overline{B}\rangle} = \left|\check{h}_{A_1}(1)\right|^2. \eea

\noindent Statistical fluctuations in the numerator and
denominator are highly correlated and therefore cancel in the
ratio.  The normalization uncertainty in the lattice currents also
largely cancels in the ratio.  Thus, all uncertainties scale as
${\cal R}-1$ rather than as ${\cal R}$.  Making use of the heavy
quark expansions of the above double ratios, we can obtain the
three $l$'s needed to construct $h_{A_1}$ to order $1/m_Q^2$
(Eq~\ref{hA1}), one from each ratio.

\bea h_{+}(1) &=&
\eta_V\left[1-l_P\left(\frac{1}{2m_c}-\frac{1}{2m_b}\right)^2\right],
\eea \bea h_{1}(1) &=&
\eta_V\left[1-l_V\left(\frac{1}{2m_c}-\frac{1}{2m_b}\right)^2\right],
 \eea \bea \check{h}_{A_1}(1) &=&
\check{\eta}_A\left[1-l_A\left(\frac{1}{2m_c}-\frac{1}{2m_b}\right)^2\right],
 \eea

\noindent where $\eta_V$ and $\check{\eta}_A$ are short-distance
coefficients of HQET.

\section{Chiral corrections to $B\to D^*$ at zero recoil}

\begin{figure}
\includegraphics[scale=.8]{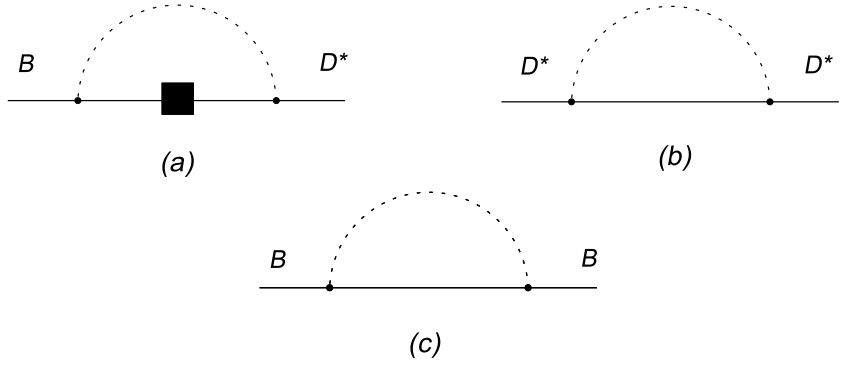}
\caption{ One-loop diagrams that contribute to $B\to D^*$.  The
solid line represents a meson containing a heavy quark, and the
dashed line represents light mesons.  The small solid circles are
strong vertices and contribute a factor of $g_\pi$.  The large
solid square is a weak interaction vertex.  Diagram (a) is a
vertex correction, and (b) and (c) correspond to wavefunction
renormalization. \label{diagrams}}
\end{figure}

The one loop diagrams that contribute to $B\to D^*$ are shown in
Fig.~\ref{diagrams}.  In general, the light mesons represented by
the dotted lines in Fig.~\ref{diagrams} include one or more
insertions of the hairpin diagrams (see \cite{aubin}) for the
singlet, axial and vector taste flavor-neutral mesons.  The
original continuum ChPT result was obtained by Randall and Wise
\cite{randall}, and was generalized to the partially quenched case
by Savage \cite{savage}. The result for the 2+1 ($m_u=m_d \neq
m_s$) full lattice QCD case including taste-breaking terms is

\bea h_{A_1}^{2+1}(1)&=&
1+\frac{X_{A}}{m_c^2}+\frac{g^2_\pi}{48\pi^2f^2}\left[\frac{1}{16}\sum_{B}(2\overline{F}_{\pi_B}+\overline{F}_{K_B})
-\frac{1}{2}\overline{F}_{\pi_I}+\frac{1}{6}\overline{F}_{\eta_I}
\right. \nonumber \\ &&
+a^2\delta'_V\left(\frac{m^2_{S_V}-m^2_{\pi_V}}{(m^2_{\eta_V}-m^2_{\pi_V})(m^2_{\pi_V}-m^2_{\eta'_V})}\overline{F}_{\pi_V}
  +\frac{m^2_{\eta_V}-m^2_{S_V}}{(m^2_{\eta_V}-m^2_{\eta'_V})(m^2_{\eta_V}-m^2_{\pi_V})}\overline{F}_{\eta_V}
 \right. \nonumber
\\ && \left. \left.
+\frac{m^2_{S_V}-m^2_{\eta'_V}}{(m^2_{\eta_V}-m^2_{\eta'_V})(m^2_{\eta'_V}-m^2_{\pi_V})}\overline{F}_{\eta'_V}\right)
+(V\rightarrow A) \right], \nonumber \\ && \label{fitfunc} \eea

\noindent where $\overline{F}_j\equiv
F\left(m_j,-\Delta^{(c)}/m_j\right)$, and

\bea F\left(m_j,x\right) &=& \frac{m_j^2}{x}\left\{x^3
\ln\frac{m_j^2}{\Lambda^2}-\frac{2}{3}x^3 -4x+2\pi \right. \nonumber \\
&& \left.
-\sqrt{x^2-1}(x^2+2)\left(\ln\left[1-2x(x-\sqrt{x^2-1})\right]-i\pi\right)\right\}
\nonumber
\\ && \longrightarrow (\Delta^{(c)})^2\ln\left(\frac{m_j^2}{\Lambda^2}\right)+{\cal O}[(\Delta^{(c)})^3], \eea

\noindent with $g_\pi$ the $D^* -D- \pi$ coupling,
$\Delta^{(c)}=m_{D^*}-m_D=142$ MeV and $X_A$ is a term that is
independent of the light quark masses that must exactly cancel the
scale dependence of the logarithms.  In principle, this term also
contains taste-breaking contributions which vanish as the lattice
spacing goes to zero. From the discussion after Eq~(\ref{mcsplit})
we see that the $D-D^*$ splitting begins at $1/m_c$, so that the
one-loop chiral corrections to the above formula begin at
$1/m_c^2$. We do not account for the smaller deviations of the
$b$-quark mass from the heavy quark limit in this calculation.

\begin{figure}
\includegraphics[scale=.9]{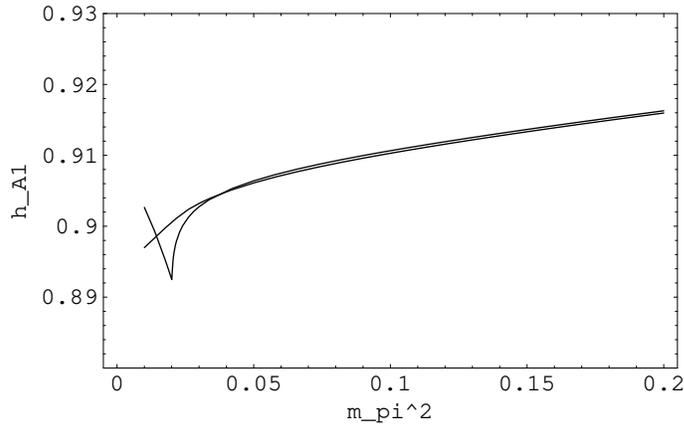}
\caption{These curves add to linear behavior the contribution from
chiral logs with $g_\pi=0.60$. The curve with the large cusp is
the continuum extrapolation; the one without the cusp includes
also the staggered effects.\label{fit}}
\end{figure}

Fig.~\ref{fit} is a plot of $h_{A_1}(1)$ vs $m_\pi^2$,
illustrating the importance of accounting for staggered effects in
the chiral limit. The part of the graph that asymptotes to a
straight line is a guess (based on the earlier quenched result
\cite{hashimoto}) as to what a linear fit to data points might
look like for the MILC lattices for $h_{A_1}(1)$.  The curves add
to the linear behavior the contribution from the chiral logs with
$g_\pi=0.60$.  The curve with the large cusp is the continuum
extrapolated curve; the one without the cusp also includes the
staggered effects with values determined from the MILC coarse
lattices ($a=0.125$ fm). The way the procedure for the
extrapolation would work in practice is one would fit lattice data
to Eq.~(\ref{fitfunc}), and then the taste-breaking effects would
be eliminated by setting the terms proportional to $a^2$ in
Eq.~(\ref{fitfunc}) to zero.  The staggered data are expected to
be linear, even when the continuum result is not; this is a
characteristic effect of taste-breaking terms due to the
mass-splittings of the different taste pions. However, simulations
would not likely be sensitive to the cusp anytime soon even if
staggering did not smooth it out, given that the cusp only occurs
at values very close to the physical pion mass.  In this case one
is especially dependent on the ChPT in order to extrapolate to the
physical light quark masses.

\end{document}